# High-Resolution X-Ray Reflectivity Study of Thin Layered Pt-Electrodes for Integrated Ferroelectric Devices

M. Aspelmeyer[1], U. Klemradt[1], W. Hartner[2], H. Bachhofer[2] and G. Schindler[2]

[1] *Sektion Physik and Center for Nano Science (CeNS), LMU München, Geschw.-Scholl-Pl. 1, 80539 München, Germany*

[2] *Infineon Technologies AG, Abt. MP TF, Otto-Hahn-Ring 6, 81739 München, Germany*

**ABSTRACT**

The structural interface properties of layered Pt/Ti/SiO$_2$/Si electrodes have been investigated using high-resolution specular and diffuse x-ray reflectivity under grazing angles. Currently this multilayer system represents a technological standard as bottom electrodes for ferroelectric thin film applications. For the electronic and ferroelectric properties of integrated devices, the film-electrode interface is of crucial importance. We focused on Pt-100nm/Ti-10nm/SiO$_2$/Si electrodes prepared under annealing conditions as employed in industrial processing, prior to the deposition of ferroelectric films. The comparison between annealed and non-annealed electrodes clearly revealed strong interfacial effects due to interdiffusion and oxidation of Ti, especially at the Pt-Ti interface. Migration of Ti into the Pt-layer results in a clear shift of the critical angle due to enclosure of TiO$_{2-x}$ within the Pt-layer. The heterogeneous distribution of TiO$_{2-x}$ suggests a diffusion mechanism mainly along the Pt-grain boundaries. At the SiO$_2$ interface a relatively weakly oxidized, remaining Ti-layer of 20 Å could be found, which is most probably correlated with the remaining adhesion to the substrate.

## INTRODUCTION

For current research on ferroelectric thin film applications such as high-ε dielectrics and non-volatile memories [1], the layered system Pt/Ti/SiO$_2$/Si is widely used as a standard bottom electrode. It combines the stability of highly conducting Pt in an oxidizing atmosphere with the good adhesion of Ti to SiO$_2$ [2]. However, the properties of integrated ferroelectrics are strongly affected by surface and interface effects, in particular by the electronic and geometric (morphological) real structure of the ferroelectric / electrode interface [3-5]. Since processing temperatures encountered during the synthesis of ferroelectric oxide films are usually above 600°C, the interfacial properties are determined by the thermal stability of the Pt-Ti bilayer. Thermal treatment in an oxygen atmosphere results in Ti oxidation accompanied by the migration of TiO$_{2-x}$ into the Pt-layer [6,7]. The extent of Ti diffusion strongly depends on bilayer properties like the Ti film thickness, the Pt microstructure and the Pt-Ti interface [8], which therefore all have to be carefully optimized for ferroelectric film integration. For example, thick Ti-layers (about 100 nm) cause the formation of a TiO$_2$ - layer on the Pt-surface [8], which can be completely incorporated into the ferroelectric film as in the case of PbZrTiO$_3$ (PZT) [9]. In the case of SrBiTa$_2$O$_9$ (SBT), a contamination of the Pt electrode with Ti significantly deteriorates the electrical properties of the ferroelectric film [10]. The latter can be prevented using thin Ti-layers (10 nm) as no additional TiO$_2$ surface layer is formed during annealing [8], or by using preoxidized TiO$_x$-layers where no interdiffusion of Ti and Pt occurs during annealing [10].

In the present work we investigate Pt-100nm/Ti-10nm/SiO$_2$/Si electrodes prepared under different annealing conditions, using high-resolution specular and diffuse x-ray reflectivity as an advanced technique for the characterization of thin films and interfaces at the nanometer scale.

**EXPERIMENT**

The Pt/Ti bilayer was prepared on a 625 nm $SiO_2$ layer, which was thermally grown on a Si substrate. After deposition of 10nm Ti and 100nm Pt films by dc magnetron sputtering, some of the electrodes were annealed for 30 min. at 650°C in $O_2$, which corresponds to the standard treatment used for ferroelectric film deposition.

Both types of electrodes were studied by specular and diffuse x-ray reflectivity under grazing angles using 9.5 keV synchrotron radiation from Hasylab (DESY) at the triple-axis spectrometer D4. For each sample, reflectivity scans (incident angle $\alpha_i$ = exit angle $\alpha_f$), offset scans ($\alpha_i - \alpha_f$ = const.), rocking scans ($\alpha_i + \alpha_f$ = const.) and detector scans ($\alpha_i$ = const.) were performed. Since the momentum transfer takes place in the scattering plane with components both perpendicular *and* parallel to the surface, these scans provide insight into lateral and vertical sample characteristics. The offset scans were done close to the specular condition ($\alpha_i - \alpha_f = \pm 0.05°$) in order to correct the reflectivity data for intensity contributions from diffuse scattering beneath the specular peak and to determine a roughness exponent H for the Pt surface (see below). The fits to the reflectivity data were performed using the Parratt-algorithm [11] and taking into account a finite resolution of $\Delta q_z = 9 \cdot 10^{-4}$ Å$^{-1}$ in vertical momentum transfer, which was determined using the primary beam divergence of 0.12 mrad and the geometric resolution function of our experimental setup [12].

**RESULTS AND DISCUSSION**

**A. SPECULAR X-RAY REFLECTIVITY**

Figure 1 shows the corrected x-ray reflectivity from the as-deposited electrode system together with the best (least square) fit to the data. The specular signal covers more than seven orders of magnitude and intensity oscillations are present over a large range of vertical momentum transfers $q_z$, indicating the presence of well defined interfaces within the layered structure. Indeed, the quantitative analysis reveals a small root-mean-square roughness $\sigma$ of about 1 nm at every interface. An overview of the fitting results is given in table 1. It should be noted that an accurate fit can only be obtained if an additional very thin surface layer is assumed for both types of electrodes. The unusual ratio of almost one between the thickness and the r.m.s.-roughness points to a gradual decrease in electron density on the length scale of the additional layer, which is most probably due to the granular structure of the free Pt surface.

The thicknesses of the Pt- and the Ti-layer agree with their nominal values within ca. 10% (113.5 nm for Pt and 9.3 nm for Ti, respectively), whereas the thickness of the $SiO_2$-layer is not resolvable due to the absorption-limited information depth of the x-ray beam. The Pt electron density derived from the fit (via the index of refraction $n = 1 - \boldsymbol{d} + i\boldsymbol{b}$) corresponds to the bulk value within 0.4%, however, the electron density of the Ti film exceeds the bulk value by 6.9%.

After annealing, the reflectivity differs significantly from that of the non-annealed sample, as can be seen in Figure 2. The critical angle is shifted to lower values by 0.008°, indicating an effective increase in the real part of the index of refraction and, therefore, a decrease in the electron density of the top layer. This correlates qualitatively with the observed increase of the Pt-layer thickness by 9.1%, resulting in a reduced electron density. However, the density reduction of the Pt-layer is found from the fit to be only 3.2 % instead of 9.1% as expected from the layer expansion (Table 1). Therefore a second factor has to be taken into account, namely the diffusion and oxidation of Ti in the Pt-layer. The polycrystalline nature of the Pt-layer provides relatively open diffusion channels, which consequently give rise to heterogeneous enclosures of $TiO_{2-x}$ in the top layer. This result is corroborated by TEM analysis performed with annealed bilayers (inset in Figure 2).

Assuming that the diffusion of Pt into the Ti layer is negligible, the dispersive and the absorptive part of the index of refraction for the mixed layer should simply be given by a sum of two contributions: one from Pt (treated as the non-annealed Pt layer but corrected for the layer expansion), and one from TiO$_{2-x}$, which is weighted according to its volume fraction $k$ within the expanded top layer, i.e.

$$\delta_{mix} = \frac{d_1}{d_2} \cdot \delta_{Pt} + k \cdot \delta_{TiO(2-x)} \qquad (1)$$

$$\beta_{mix} = \frac{d_1}{d_2} \cdot \beta_{Pt} + k \cdot \beta_{TiO(2-x)} \qquad (2)$$

where $d_1$ and $d_2$ are the thickness of the top layer before and after the anneal, and $1-\delta$ and $\beta$ are the real and imaginary part of the index of refraction. $\delta_{Pt}$ and $\beta_{Pt}$ refer to the non-annealed state of the sample, and $\delta_{mix}$ and $\beta_{mix}$ are taken from the fit results for the mixed Pt-TiO$_{2-x}$ layer of the annealed sample. In a first approximation, the values for $\delta_{TiO(2-x)}$ and $\beta_{TiO(2-x)}$ are taken from the TiO$_2$ layer (see table 1). This formalism covers both limiting cases, as $\delta_{mix} = (d_1/d_2) \cdot \delta_{Pt}$ for $k = 0$ and $\delta_{mix} = \delta_{TiO2-x}$ for $k = 1$, since $(d_1/d_2)$ approaches zero if, as was assumed, the amount of Pt in the layer is kept constant. Note also that all variables in equations (1) and (2) are determined by the experiment.

From the fits to the reflectivity data the values for δ can be determined quite precisely within a few percent, whereas the values for β can have errors of up to 100% owing to the fact that the specular reflectivity is in general not very sensitive to the absorption in deeply buried layers. Usually, in systems of known compositions this problem can be circumvented by keeping the ratio δ/β constant. In the case of intermixing this constraint is missing, and δ and β have to be decoupled for fitting the data. Consequently, the volume fraction κ should be determined by equation (1), which yields $k = (29 \pm 3)$ %. In other words, this result indicates that ca. 30% of the volume of the expanded former Pt-layer is occupied by migrated and oxidized Ti ($TiO_{2-x}$). This value is surprisingly large compared to the value which can be calculated from the (bulk) densities of Ti and $TiO_2$ [13], resulting in $\kappa = 15\%$. However, it should be kept in mind that our fitting procedure does not take into account the spatial variation of $TiO_{2-x}$ perpendicular to the interfaces, but rather assumes a homogeneous distribution inside the film. Although it is in principle possible to introduce a graded mixed layer, we chose not to introduce additional parameters in our structural model, since already the simple homogeneous-layer model gives essentially a perfect fit. As systematic errors are hard to estimate for the determination of $\kappa$ both from x-rays and the bulk densities, the numerical values should be regarded as assessments.

Figure 3 shows the electron density profiles of the samples as were calculated from fits to the reflectivity data. After annealing the separation of two different regions within the Ti-layer is clearly discernible (Figure 3). Just below the top (intermixed) layer we find a 9.2 nm thick layer with strongly reduced electron density compared to Ti (by almost 40%). Its density is still enhanced compared to $TiO_2$ (by 35%), suggesting an incompletely oxidized layer of $TiO_{2-x}$. Also, an additional 2.6 nm thick layer with an electron density reduced by only 12 % (compared to bulk Ti) has to be assumed. It is well separated from the above $TiO_{2-x}$ by an extremely sharp interface ($\sigma = 0.5$ nm); its electron density points to a remaining Ti-layer which is clearly less oxidized than the above mentioned $TiO_{2-x}$-layer. Although annealing temperatures of about 650°C cause oxygen to diffuse out of the $SiO_2$-layer due to $SiO_2$ dissociation [14], the dominating oxidation process takes place at the Pt interface and within the Pt-layer. Since solubility and diffusion rate of oxygen in bulk Pt are very low [15], the oxygen mainly propagates along the Pt grain boundaries.

Despite of the sharp interface within the oxidized Ti-layer, the interface to the neighboring $SiO_2$-layer shows a strong enhancement in r.m.s.-roughness by 120% compared to the non-annealed sample. The oxidation and interdiffusion of Ti and the extremely low thickness of the remaining Ti-layer could be responsible for the loss of adhesion of the annealed electrode system to the $SiO_2$ substrate as was reported previously [8, 21]. The observed increase in thickness of several percent for both the Pt and the Ti-layers points to a relaxation process during annealing. This correlates with previous experiments that indicate an increase in tensile stress due to the annealing procedure [10].

## B. DIFFUSE X-RAY REFLECTIVITY

Compared to the changes in the reflectivity the diffuse intensity of the rocking scans does not show significant differences for the annealed and the non-annealed case, indicating the dominance of interdiffusion between the layers. Furthermore, in both offset scans for the non-annealed and the annealed case periodic oscillations of intensity are absent, indicating the lack of vertical correlation of interfacial roughness between different layers, especially between Pt and Ti (Figure 4). The minute changes in the absolute scattered intensity can be attributed to a changed lateral correlation length of the top (Pt) layer. Since the Pt-layer thickness is relatively large compared to the penetration depth of the x-rays, changes in the diffuse intensity mainly stem from changes in the height-height-correlation-function of the uppermost surface, i.e. the morphology of the Pt-surface. For calculations of the diffuse intensity distribution, the electrode system was therefore simplified to a single Pt-layer on $SiO_2$-substrate. Simulations of the rocking scans and detector scans have been performed using an algorithm provided by Holý [16] [17] which is based on the DWBA-formulation of the diffuse scattering problem for grazing incident angles [18].

Assuming self-affinity for the surface roughness, the height-height-correlation function C(R) can be described by $C(R) = \sigma^2 \exp\{-(R/\xi)^{2H}\}$, where $\sigma$ denotes the r.m.s.- roughness of the surface, $\xi$ is the lateral correlation length parallel to the surface and H is the roughness exponent corresponding to the jaggedness of the surface. H can be determined by analyzing the diffuse intensity that accompanies the specular peak, which is basically the information content of the offset scans that have been performed. For small offsets close to the specular peak the diffuse intensity decays with increasing vertical momentum transfer $q_z$ following the power law

$$I(q_z) \propto q_z^{(3+1/H)} \qquad (4)$$

provided that the slit settings of the experimental setup result in integration of the diffuse signal perpendicular to the scattering plane and provided that the angles of incidence are chosen large enough to overcome the full illumination of the sample [19]. For both samples the fit to the offset scans reveals a roughness exponent $H = 0.95 \pm 0.1$ (Figure 4a). Therefore, using $H = 1$, the decay of the diffuse intensity in the offset scan could also be simulated (Figure 4). The fits to the rocking- and detector scans show an increase in the lateral correlation length by almost 50% from $\xi = 520$ Å before the anneal to $\xi = 820$ Å after annealing the sample, indicating a change of the Pt-surface morphology on a nanometer scale due to the annealing procedure. These results correlate with an observed increase of grain size in the Pt-layer from 500 Å up to 1000 Å on annealing the sample [20].

**CONCLUSIONS**

X-ray specular and diffuse reflectivity studies have been performed on layered Pt/Ti/SiO$_2$/Si electrode systems prepared under different annealing conditions prior to ferroelectric film deposition. After annealing the electrodes for 30 min at 650°C, drastic interdiffusion effects are clearly revealed by the reflectivity data. Both the Pt- and the Ti-layer are found to expand vertically. Strong depletion of Ti not only takes place close to the Pt-layer but Ti also oxidizes and migrates into the Pt film, where $TiO_{2-x}$ is found to occupy a volume fraction of ca. 30%. This suggests an additional non-homogeneous variation of $TiO_{2-x}$ within the mixed layer, although the reflectivity data is already accurately fitted by a homogeneous distribution of $TiO_{2-x}$. At the $SiO_2$ interface we find a 25 Å layer of relatively weakly oxidized Ti, which could be responsible for the remaining adhesion properties on the $SiO_2$ substrate. The signal of the diffuse scattered intensity basically contains information of the Pt-surface morphology and is in agreement with ex-situ microscopy-data taken from those surfaces. In a next step, electrodes in their combination with ferroelectric top layers will be studied with regard to influences of the interfacial structure on the ferroelectric properties of the material.

## ACKNOWLEDGEMENT

We acknowledge the excellent beamline support at D4 (Hasylab) by Dr. Dimitri Novikov, Dr. Hermann Franz and Rüdiger Novak. Also we would like to thank Christian Braun (HMI Berlin) for providing a software implementation of the Parratt-algorithm (parrat32, Version 1.5.2.), Prof. Dr Vaclav Holý (Masaryk University, Brno) for the Fortran-code to simulate diffuse scattering, Dr. Hans Cerra (Siemens AG, Corporate Research) for TEM analysis and Dr. Michael P. Vogel (MPI Golm) for fruitful discussion.

**FIGURE CAPTIONS**

[Figure 1] Specular x-ray reflectivity from the non-annealed sample (circles: experimental data, line: best fit). The rocking scan in the inset was performed at an exit angle of 0.2° (marked by the arrow). The extremely low half-width indicates an excellent sample flatness indispensable for quantitative high-resolution studies.

[Figure 2] Annealing in $O_2$-atmosphere strongly affects the specular x-ray reflectivity (circles: as-deposited, triangles: annealed). The shift in the critical angle (inset) can be explained by an expansion of the Pt film together with the oxidation of Ti migrated into the film. The TEM-picture shows the heterogeneous incapsulation of $TiO_{2-x}$ within the Pt-layer resulting in averaged x-ray optical properties of the layer.

[Figure 3] Electron density profiles calculated from the best fits to the specular x-ray reflectivity. The position of the mean free Pt surface was defined to be zero in both cases. The horizontal lines indicate the tabulated standard values for the given materials. For clarity reasons the plot does not show the full extension of the 100nm top layer.

[Figure 4] (a) Detector scans from the non-annealed sample at an incident angle of 1.0° (circles) and from the annealed sample at an incident angle of 3° (squares). Both curves were simulated using a roughness exponent of $H = 1$ derived from the decay of the offset-scan intensity with $q_z$ (inset). The simulation (lines) yields $\xi = 520$ Å and $\sigma = 10$ Å for the non-annealed sample and $\xi = 820$ Å and $\sigma = 11$ Å for the annealed sample.

(b) Simulation of the rocking scans using the fit results from above.

[Table 1] Results from the least square fits to the x-ray reflectivity curves. The fitting parameters are layer thickness d, interfacial root-mean-square roughness $\sigma$ and the real and imaginary part of the index of refraction $\delta$ and $\beta$, respectively. Fixed parameters are given in squared brackets.

|  |  | design values | as-deposited | annealed |
|---|---|---|---|---|
| **surface layer** | d [Å] | - | 10.6 ± 0.5 | 8.4 ± 0.4 |
|  | σ [Å] | - | 12.4 ± 0.1 | 17.3 ± 0.1 |
|  | δ | - | (3.12 ± 0.08)·$10^{-5}$ | (3.35 ± 0.03)·$10^{-5}$ |
|  | β | - | (1.01 ± 0.05)·$10^{-5}$ | (0.47 ± 0.03)·$10^{-5}$ |
| Pt / TiO$_{2-x}$ | d [Å] | - | - | 1238.8 ± 2.5 |
|  | σ [Å] | - | - | 8.9 ± 0.2 |
|  | δ | - | - | (3.57 ± 0.01)·$10^{-5}$ |
|  | β | - | - | (0.27 ± 0.02)·$10^{-5}$ |
| Pt | d [Å] | 1000 | 1135.2 ± 2.5 | - |
|  | σ [Å] | - | 11.5 ± 0.3 | - |
|  | δ | 3.647·$10^{-5}$ | (3.69 ± 0.03)·$10^{-5}$ | - |
|  | β | 0.248·$10^{-5}$ | (0.28 ± 0.01)·$10^{-5}$ | - |
| **TiO$_2$** | d [Å] | - | - | 92 ± 2 |
|  | σ [Å] | - | - | ³ 18 |
|  | δ | 0.487·$10^{-5}$ | - | (0.66 ± 0.03)·$10^{-5}$ |
|  | β | 1.830·$10^{-7}$ | - | (2.5 ± 2)·$10^{-7}$ |
| **Ti** | d [Å] | 100 | 93 ± 2 | 26 ± 1 |
|  | σ [Å] | - | 16 ± 2 | 5.0 ± 0.5 |
|  | δ | 0.967·$10^{-5}$ | (1.03 ± 0.03)·$10^{-5}$ | (0.85 ± 0.03)·$10^{-5}$ |
|  | β | 6.042·$10^{-7}$ | (6.5 ± 5)·$10^{-7}$ | (5.3 ± 5)·$10^{-7}$ |
| **SiO$_2$** | d [Å] | 6250 | [6250] | [6250] |
|  | σ [Å] | - | 4.0 ± 0.1 | 9 ± 1 |
|  | δ | 5.116·$10^{-6}$ | (4.9 ± 0.2)·$10^{-6}$ | (4.9 ± 0.8)·$10^{-6}$ |
|  | β | 4.769·$10^{-8}$ | (4.8 ± 4)·$10^{-8}$ | (4.8 ± 4)·$10^{-8}$ |

Table 1 / Aspelmeyer et al.

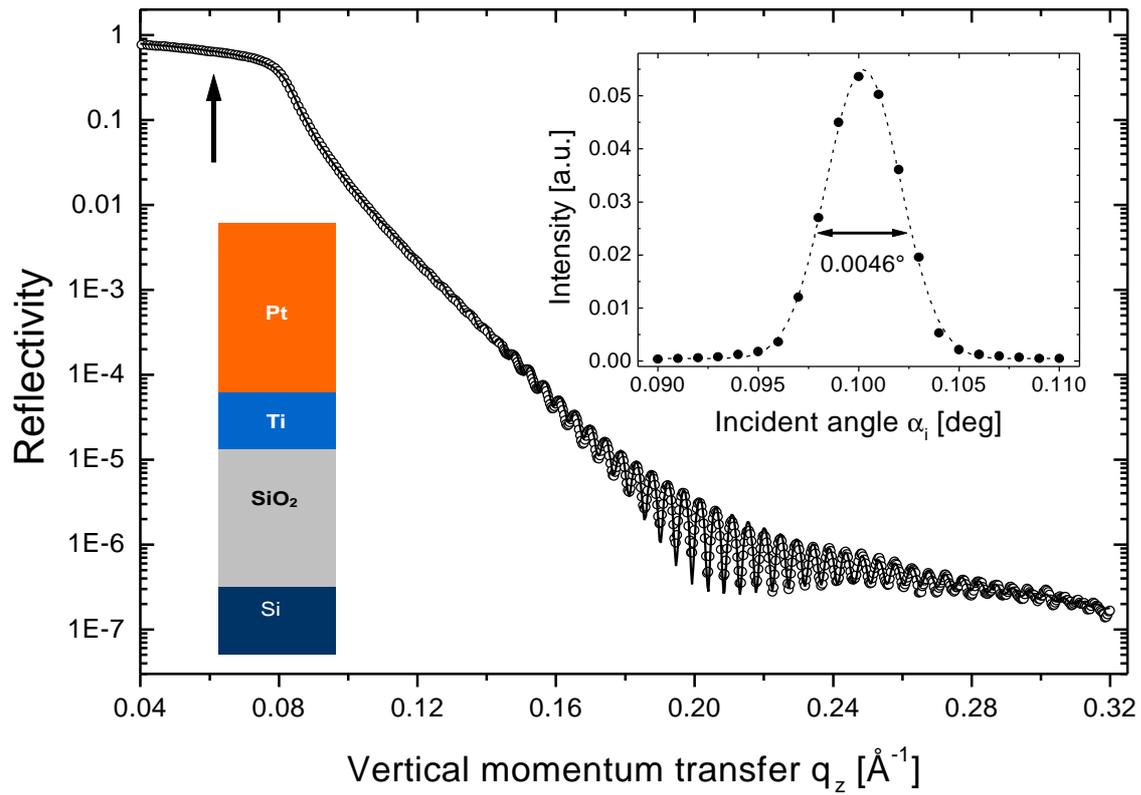

Figure 1 / Aspelmeyer et al.

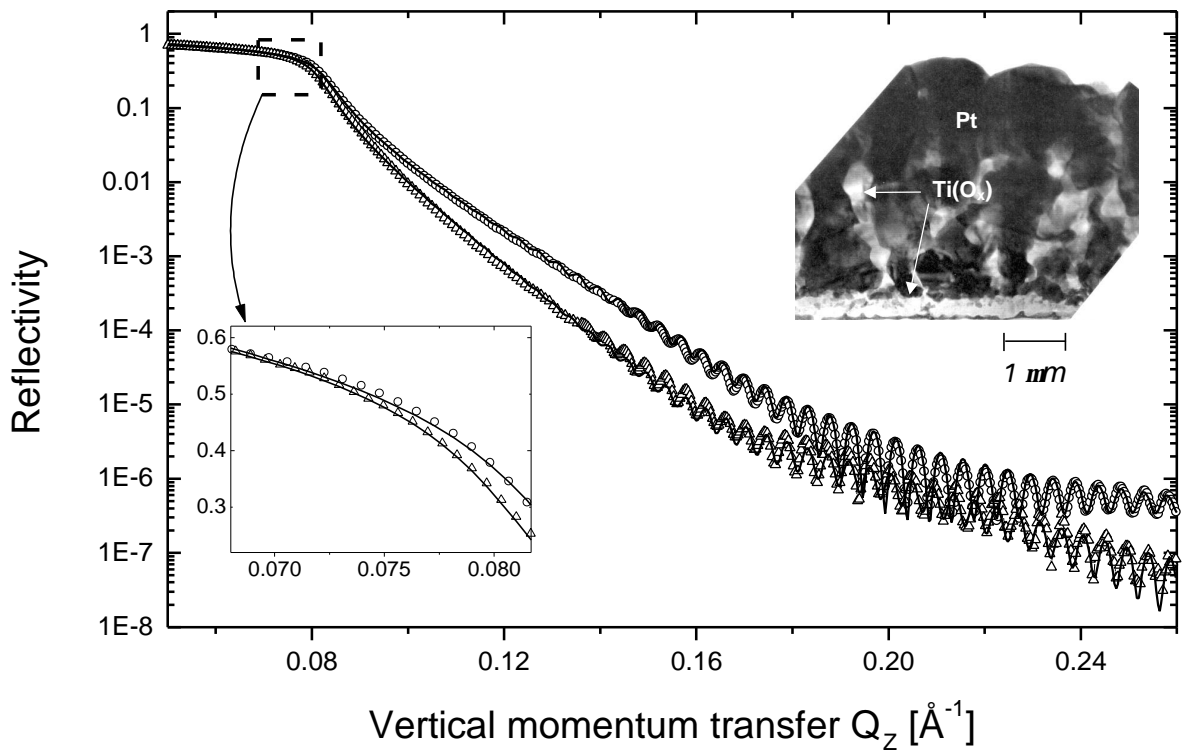

Figure 2 / Aspelmeyer et al.

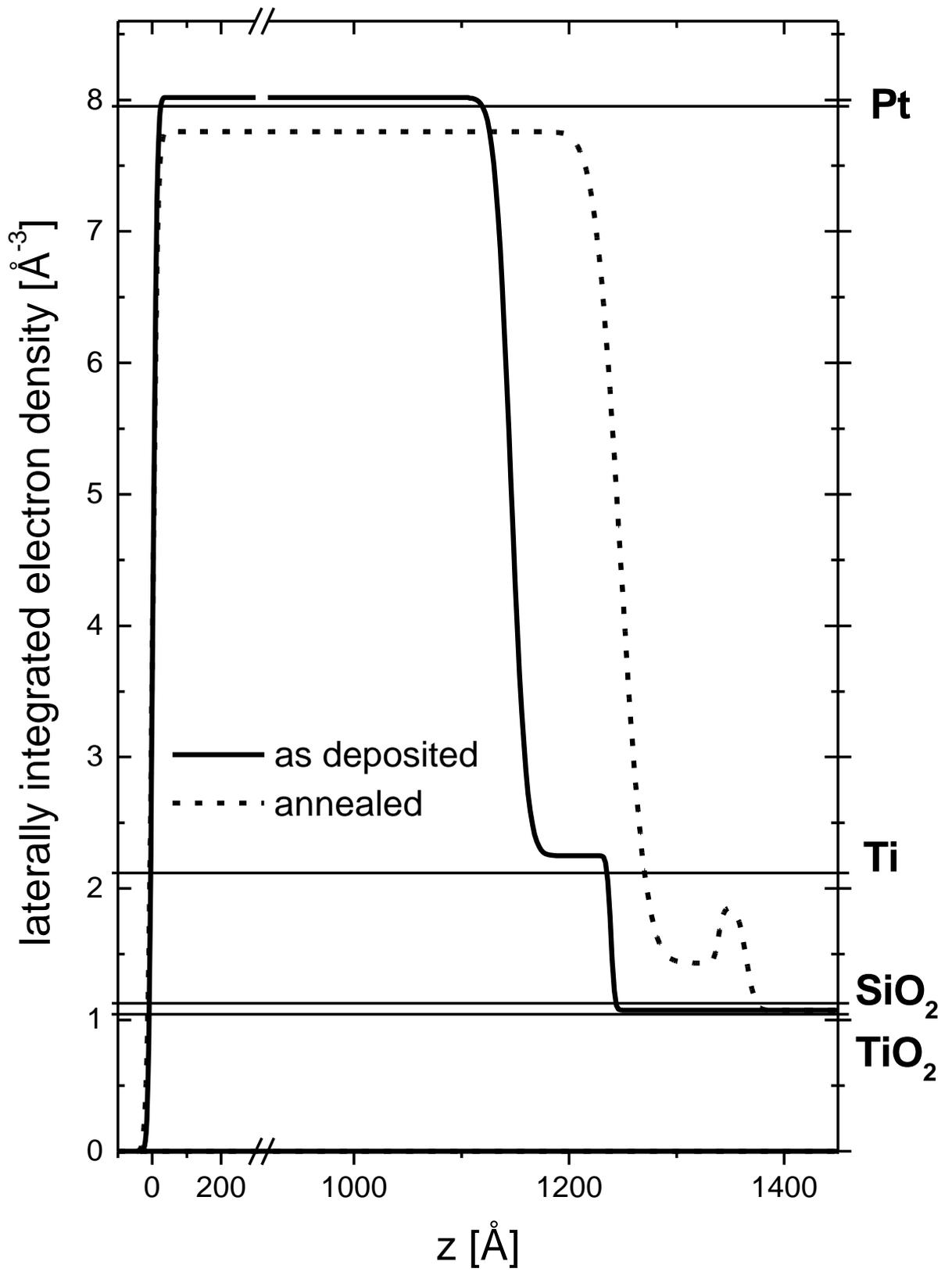

Figure 3 / Aspelmeyer et al.

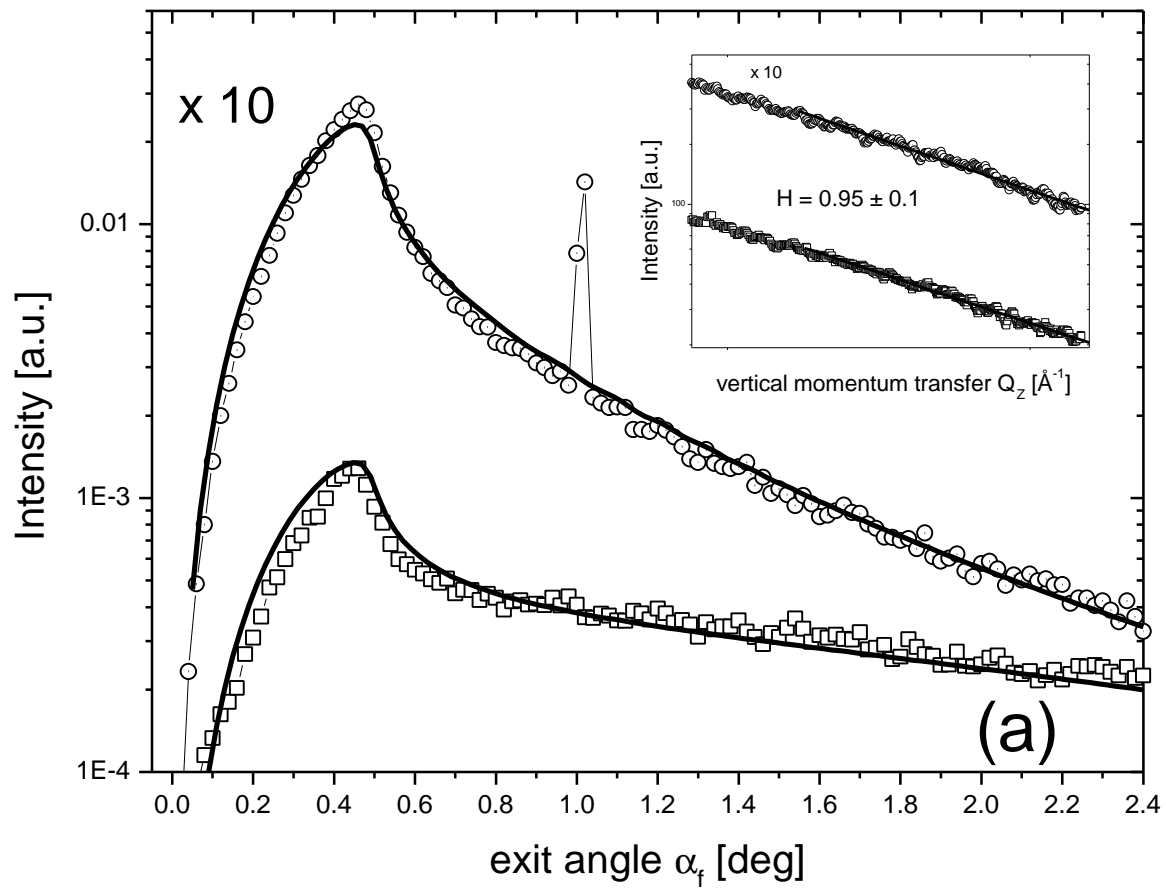

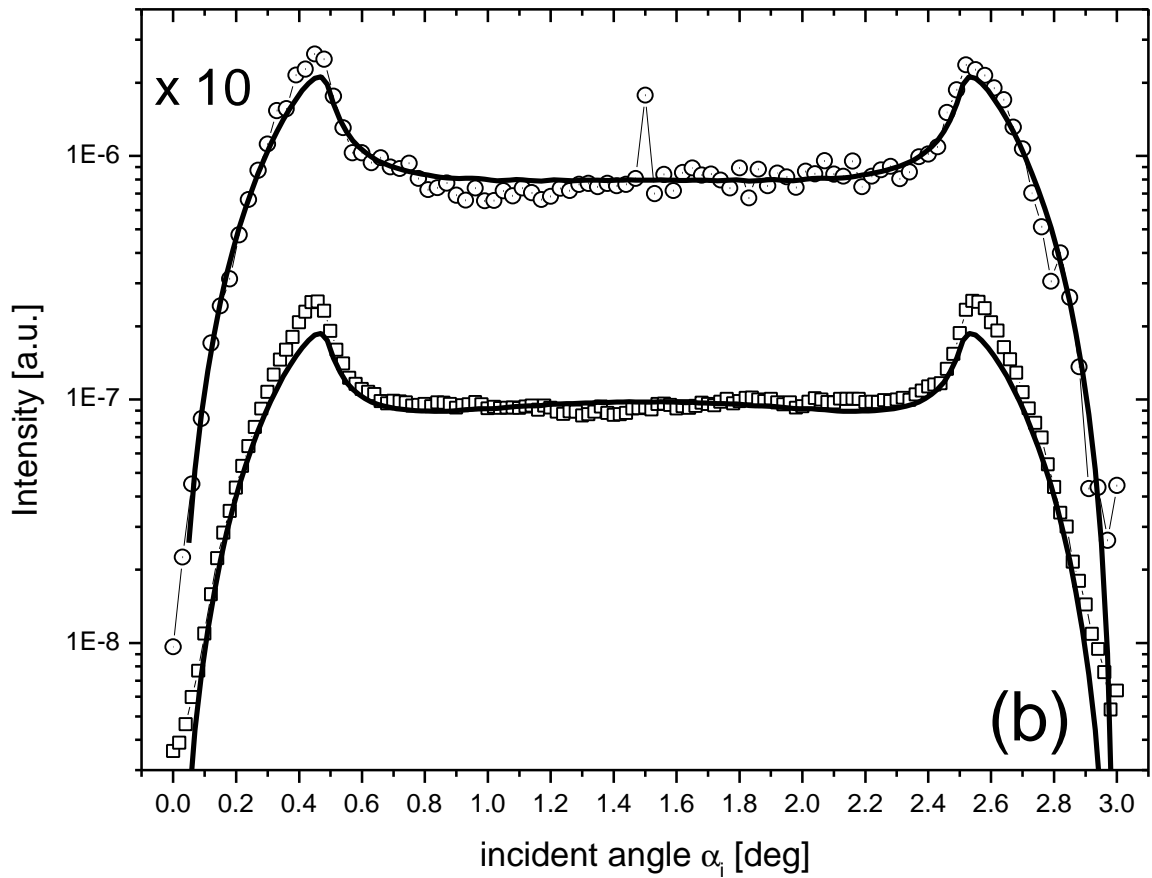

Figure 4 / Aspelmeyer et al.

15